\documentclass{aip-cp}

\usepackage[numbers]{natbib}
\usepackage{rotating}
\usepackage{epsf}
\usepackage{amssymb}
\usepackage{amsfonts}
\usepackage{psfrag,epsfig,graphicx}

\newcommand{\be}{\begin{eqnarray}}
\newcommand{\ee}{\end{eqnarray}}
\newcommand{\ba}{\begin{array}}
\newcommand{\ea}{\end{array}}
\newcommand{\bi}{\begin{itemize}}
\newcommand{\ei}{\end{itemize}}

\def\beq{\begin{equation}}
\def\eeq{\end{equation}}
\newcommand{\eq}{\end{equation}}
\def\bea{\begin{eqnarray}}
\def\beqa{\begin{eqnarray}}
\def\eea{\end{eqnarray}}
\def\eqa{\end{eqnarray}}

\def\dv{\vec{\Delta}_t}
\def\ar{\alpha_\rho}

\begin{document}

\title{Accessing Generalized Parton Distributions in Exclusive Photoproduction of a $\gamma \rho$ Pair with a Large Invariant Mass}

\author[aff1]{R.~Boussarie\corref{cor1}}
\author[aff2]{B.~Pire\corref{cor2}}
\author[aff3]{L.~Szymanowski\corref{cor3}}
\author[aff4]{S.~Wallon\corref{cor4}}

\affil[aff1]{Institute of Nuclear Physics, Polish Academy of Sciences, Radzikowskiego 152, PL-31-342 Krak\'ow, Poland}
\affil[aff2]{Centre de Physique Th\'eorique, \'{E}cole Polytechnique, CNRS, \\
Universit\'e Paris-Saclay, 91128 Palaiseau, France}
\affil[aff3]{National Centre for Nuclear Research (NCBJ), Ho\.za 69, 00-681 Warsaw, Poland}
\affil[aff4]{Laboratoire de Physique Th\'{e}orique (UMR 8627), CNRS, Univ. Paris-Sud, \\
Universit\'{e} Paris-Saclay, 91405 Orsay Cedex, France \\ UPMC, Universit\'{e} Paris 06, Facult\'{e} de Physique, 4 place Jussieu, 75252 Paris, France \\}
\corresp[cor1]{renaud.boussarie@ifj.edu.pl}
\corresp[cor2]{bernard.pire@polytechnique.edu}
\corresp[cor3]{lech.szymanowski@ncbj.gov.pl}
\corresp[cor4]{samuel.wallon@th.u-psud.fr}

\maketitle

\begin{abstract}
We propose and study the 
 photoproduction of a $\gamma\,\rho$ pair with a large invariant mass and  a small transverse momentum of  the final nucleon, as a way to access generalized parton distributions. In the kinematics of JLab~12-GeV, we demonstrate the feasibility of this measurement.
\end{abstract}

\section{INTRODUCTION}

In this contribution, we report on our calculation \cite{BoussarieProc} of the scattering amplitude for the process
\begin{equation}
\gamma(q) + N(p_1) \rightarrow \gamma(k) + \rho^0(p_\rho,\varepsilon(p_\rho)) + N'(p_2)\,.
\label{process1}
\end{equation}
%
\begin{figure}[h!]

\psfrag{TH}{$T_H$}
\psfrag{Pi}{$\pi$}
\psfrag{P1}{$\,\phi$}
\psfrag{P2}{$\,\phi$}
\psfrag{Phi}{$\,\phi$}
\psfrag{Rho}{$\rho$}
\psfrag{tp}{$t'$}
\psfrag{s}{$s$}
\psfrag{x1}{\raisebox{-.1cm}{$\hspace{-.4cm}x+\xi$}}
\psfrag{x2}{\raisebox{-.1cm}{$\!x-\xi$}}
\psfrag{RhoT}{$\rho_T$}
\psfrag{t}{$t$}
\psfrag{N}{$N$}
\psfrag{Np}{$N'$}
\psfrag{M}{$M^2_{\gamma \rho}$}
\psfrag{GPD}{$\!GPD$}

\centerline{
\raisebox{1.6cm}{\includegraphics[width=11pc]{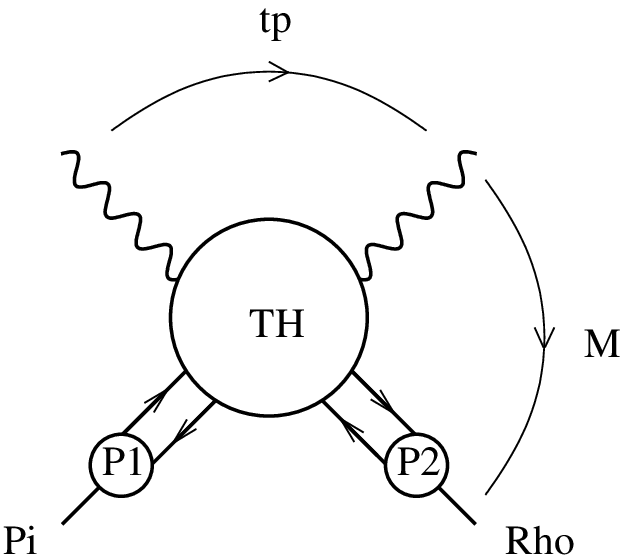}}~~~~~~~~~~~~~~
\psfrag{TH}{$\,  T_H$}
\includegraphics[width=11pc]{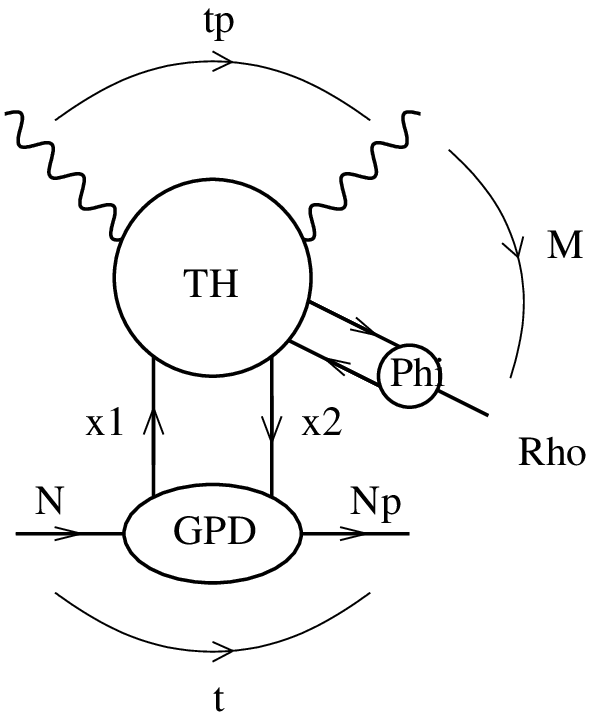}}

\caption{a) Factorization of the amplitude for the process $\gamma + \pi \rightarrow \gamma + \rho $ at large $s$ and fixed angle (i.e. fixed ratio $t'/s$); b) replacing one DA by a GPD leads to the factorization of the amplitude  for $\gamma + N \rightarrow \gamma + \rho +N'$ at large $M_{\gamma\rho}^2$\,.}
\label{Fig:feyndiag}
\end{figure}
%
Photoproduction of a pair of particles with large invariant mass 
is a natural case for using an extension of collinear QCD factorization theorems which describe the amplitudes
for deeply virtual
 Compton scattering (DVCS) and deeply virtual meson production~\cite{review}. The simplest example is timelike Compton scattering \cite{TCS} where the  produced particles are a lepton pair  with large invariant mass $Q$. In process (\ref{process1}), a wide angle Compton scattering subprocess $\gamma (q\bar q) \to \gamma \rho $ characterized by the large scale $M_{\gamma \rho}$ (the final state invariant mass) factorizes from generalized parton distributions (GPDs). This  scale $M_{\gamma \rho}$ is related to the large transverse momenta transmitted to  the final photon and to  the final meson, but we insist that the final $\gamma \rho$ pair has an overall small transverse momentum (noted $-\Delta_\perp$). At leading twist, $\rho$ meson DAs have different chirality  when the $\rho$ is transversally or longitudinally polarized. Thus,  separating the  transverse  (resp. longitudinal) polarization of the $\rho$ meson allows one to get access to  chiral-odd (resp.  chiral-even) GPDs. This opens a new way for the 
extraction of GPDs and should thus constitute a useful check of their universality.

The study of such $2\to3$ processes was initiated in Ref.~\cite{IPST-eps}, where the process under study was the high-energy diffractive photo- (or electro-) production
 of two vector mesons, the hard probe being the virtual "Pomeron" exchange (and the hard scale being the virtuality of this Pomeron). A similar strategy has also been advocated in Ref.~\cite{Beiyad:2010cxa,kumano} to enlarge the number of processes which could be used to extract information on chiral-even GPDs.

While the magnitude of chiral-even GPDs is known, chiral-odd GPDs are very poorly known. Still, we claim below that 
the experimental study of  process (\ref{process1}) should
be feasible, with a very high counting rate for $\rho_L$ and a sizable one for $\rho_T$ at JLab~12-GeV.

The now classical proof of factorization of exclusive scattering at fixed angle and large energy~\cite{LB} allows to write the leading-twist
 amplitude for the process $\gamma + \pi \rightarrow \gamma + \rho $ as the convolution of mesonic distribution amplitudes (DAs) of $\pi$ and $\rho$ and a hard scattering subprocess amplitude $\gamma  +( q + \bar q) \rightarrow \gamma + (q + \bar q) $ with the meson state replaced by a collinear quark-antiquark pair, as shown in Figure~\ref{Fig:feyndiag}a. 
Besides, the strategy used in the factorization procedure of the exclusive meson electroproduction amplitude near the forward region~\cite{fact} can be extended by replacing in Figure~\ref{Fig:feyndiag}a the lower left meson DA by a $N \to N'$ GPD, and thus get Figure~\ref{Fig:feyndiag}b. Indeed the same collinear factorization property bases the validity of the leading-twist approximation, which either replaces the meson wave function by its DA or describes the $N \to N'$ transition through GPDs. A slight difference is that light-cone fractions ($z, 1- z$) leaving the DA are positive, but the corresponding fractions ($x+\xi,\xi-x$) may be positive or negative in the case of the GPD. Our explicit calculation
 shows that this difference does not spoil the factorization property, at least at leading order.

We define
$
P^\mu = \frac{p_1^\mu + p_2^\mu}{2} ~,~ \Delta^\mu = p_2^\mu - p_1^\mu\,.
$
In the Sudakov basis  spanned by  the light-cone vectors $p$ and $n$ ($2p\cdot n = s $), 
\beqa
 p_1^\mu &=& (1+\xi)\,p^\mu + \frac{M^2}{s(1+\xi)}\,n^\mu~, \quad p_2^\mu = (1-\xi)\,p^\mu + \frac{M^2+\vec{\Delta}^2_t}{s(1-\xi)}n^\mu + \Delta^\mu_\bot\,, \quad q^\mu = n^\mu ~,\\
k^\mu &=& \alpha \, n^\mu + \frac{(\vec{p}_t-\vec\Delta_t/2)^2}{\alpha s}\,p^\mu + p_\bot^\mu -\frac{\Delta^\mu_\bot}{2}~,
 p_\rho^\mu = \alpha_\rho \, n^\mu + \frac{(\vec{p}_t+\vec\Delta_t/2)^2+m^2_\rho}{\alpha_\rho s}\,p^\mu - p_\bot^\mu-\frac{\Delta^\mu_\bot}{2}\,,\nonumber 
\eqa
where 
$M$, $m_\rho$ are the nucleon and  $\rho$ meson masses.
The total cms energy squared of the $\gamma$-N system is
$
S_{\gamma N} = (q + p_1)^2 = (1+\xi)s + M^2\,.
$
On the nucleon side, the squared transferred momentum is
\begin{equation}
\label{transfmom}
t = (p_2 - p_1)^2 = -\frac{1+\xi}{1-\xi}\vec{\Delta}_t^2 -\frac{4\xi^2M^2}{1-\xi^2}\,.
\end{equation}
The other relevant Mandelstam invariants read
\begin{eqnarray}
\label{M_pi_rho}
s'&=& ~(k +p_\rho)^2 = ~M_{\gamma\rho}^2= 2 \xi \, s \left(1 - \frac{ 2 \, \xi \, M^2}{s (1-\xi^2)}  \right) - \dv^2 \frac{1+\xi}{1-\xi}\,, \\
\label{t'}
- t'&=& -(k -q)^2 =~\frac{(\vec p_t-\vec\Delta_t/2)^2}{\alpha} \;,\\
\label{u'}
- u'&=&- (p_\rho-q)^2= ~\frac{(\vec p_t+\vec\Delta_t/2)^2+(1-\alpha_\rho)\, m_\rho^2}{\alpha_\rho}
 \; .
\end{eqnarray}
The squared invariant mass $M^2_{\gamma\rho}$ of the ($\gamma$ $\rho^0$) system
provides the hard scale.  At leading twist, the hard part is computed in the generalized Bjorken limit: neglecting $\vec\Delta_\bot$ in front of $\vec p_\bot$ as well as hadronic masses, it amounts to
\beqa
\label{skewness2}
M^2_{\gamma\rho} \approx  \frac{\vec{p}_t^2}{\alpha\bar{\alpha}} 
~;~
\ar \approx 1-\alpha =\bar \alpha~;~
\xi =2  \frac{\tau}{2-\tau} ~,\tau \approx 
\frac{M^2_{\gamma\rho}}{S_{\gamma N}-M^2} ~;
-t'  \approx  \bar\alpha\, M_{\gamma\rho}^2  ~,-u'  \approx  \alpha\, M_{\gamma\rho}^2 \,.
\eqa

\section{THE HARD AMPLITUDE}

The scattering amplitude of process (\ref{process1})  reads
\beq
\label{AmplitudeFactorized}
\mathcal{A}(t,M^2_{\gamma\rho},p_T)  =\frac{1}{\sqrt{2}} \int_{-1}^1dx\int_0^1dz\ (T^u(x,z) \, F^{u}(x,\xi,t)    -T^d(x,z)\, F^{d}(x,\xi,t)) \,\Phi_{||,\bot}(z)\,,
\eq
\def\diagici{2.cm}
\begin{figure}[h!]
\psfrag{z}{\begin{small} $z$ \end{small}}
\psfrag{zb}{\raisebox{0cm}{ \begin{small}$\bar{z}$\end{small}} }
\psfrag{gamma}{\raisebox{+.1cm}{ $\,\gamma$} }
\psfrag{pi}{$\,\pi$}
\psfrag{rho}{$\,\rho$}
\psfrag{TH}{\hspace{-0.2cm} $T_H$}
\psfrag{tp}{\raisebox{.5cm}{\begin{small}     $t'$       \end{small}}}
\psfrag{s}{\hspace{.6cm}\begin{small}$s$ \end{small}}
\psfrag{Phi}{ \hspace{-0.3cm} $\phi$}
\hspace{-.2cm}
\begin{tabular}{ccccc}
\includegraphics[width=\diagici]{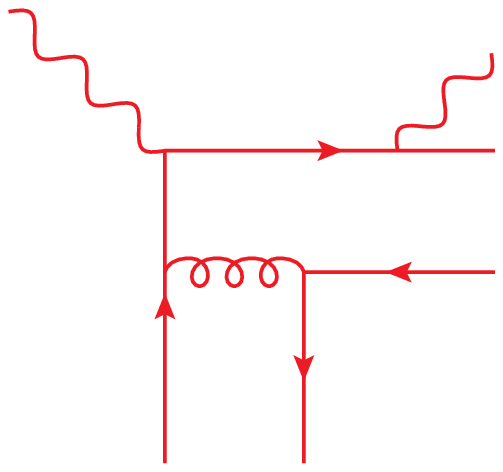}
&
\includegraphics[width=\diagici]{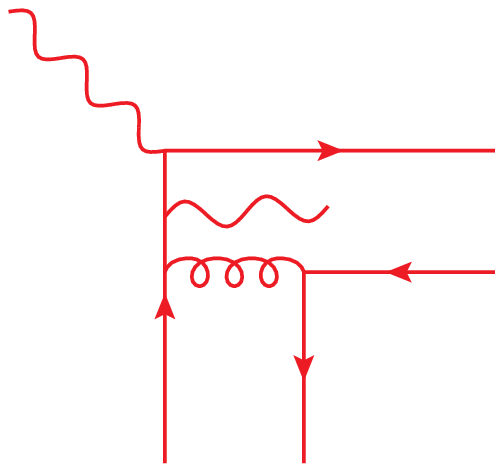}
&
\includegraphics[width=\diagici]{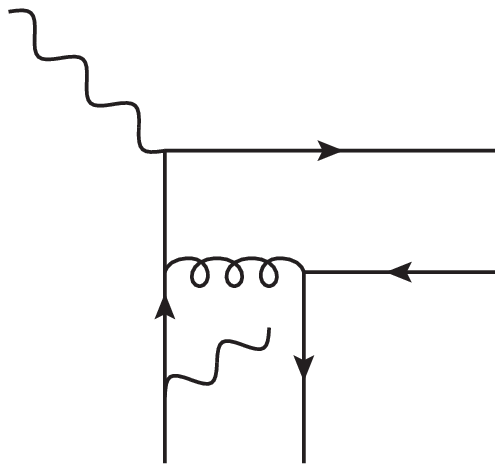}
&
\includegraphics[width=\diagici]{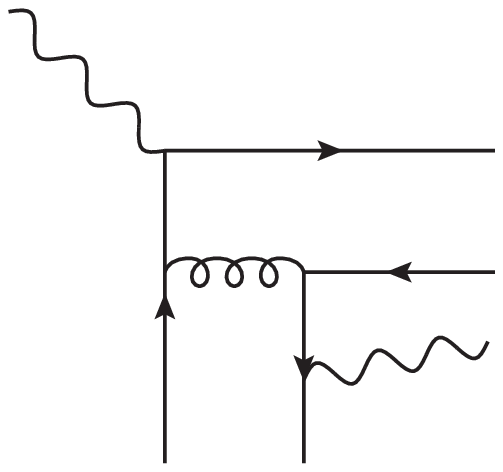}
&
\includegraphics[width=\diagici]{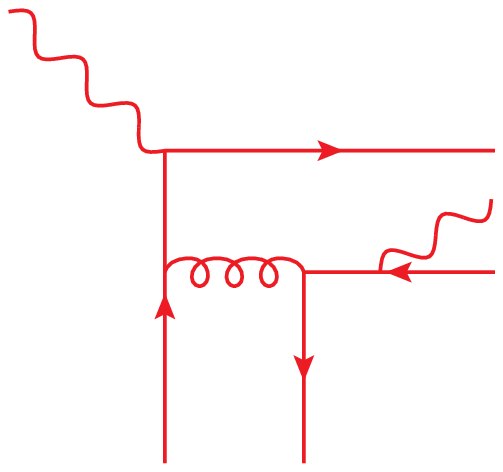} \\
$A_1$ & $A_2$ & $A_3$ & $A_4$ & $A_5$ \\
\\
\includegraphics[width=\diagici]{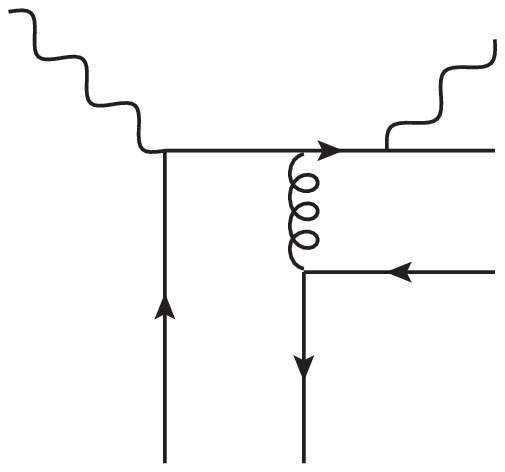}
&
\includegraphics[width=\diagici]{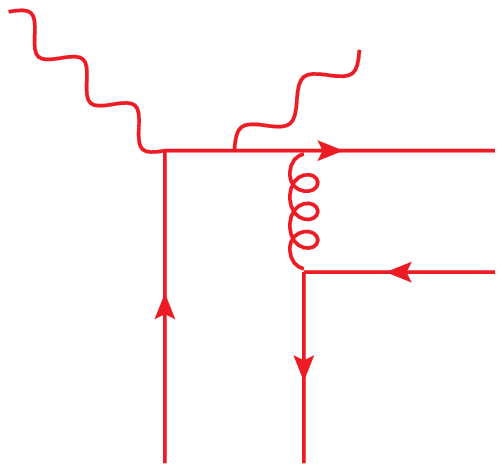}
&
\includegraphics[width=\diagici]{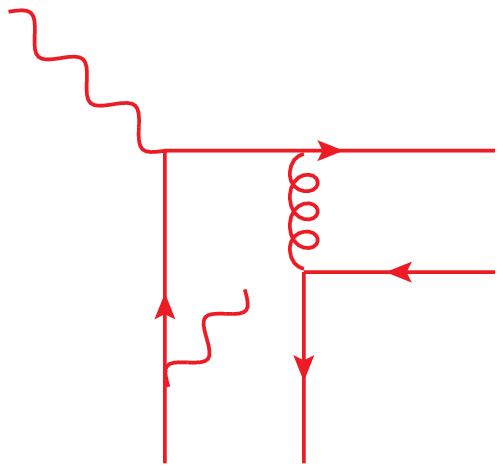}
&
\includegraphics[width=\diagici]{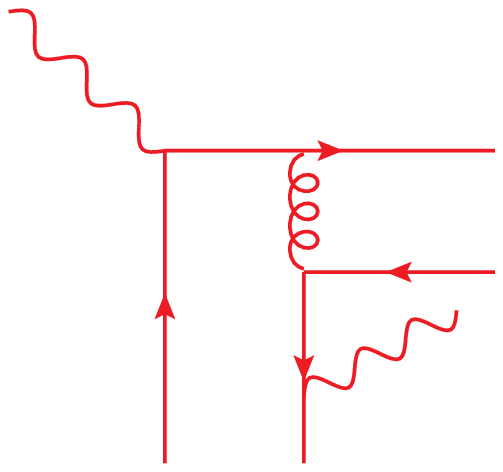}
&
\includegraphics[width=\diagici]{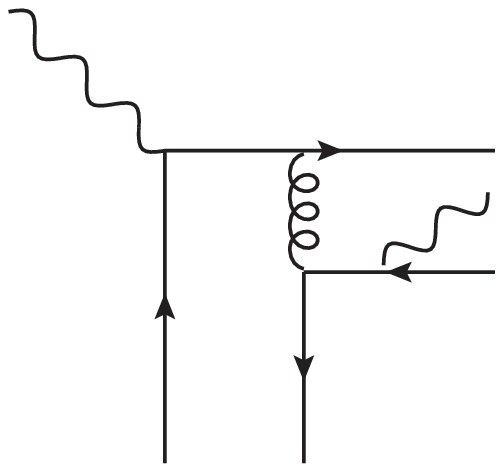}
\\
$B_1$ & $B_2$ & $B_3$ & $B_4$ & $B_5$
\end{tabular}
\caption{Half of the Feynman diagrams contributing to the hard amplitude. In the chiral-odd case, $A_3$, $A_4$ and $B_1$, $B_5$ are the only contributing diagrams (the red diagrams cancel in this case).}
\label{Fig:diagrams}
\end{figure}
%
%
where $T^u$ and $T^d$ are the hard parts of the amplitude where the photon couples respectively to a $u$-quark  and to a $d$-quark and $F^{u,d}$ is a chiral-odd or chiral-even GPD. This decomposition, with the $\frac{1}{\sqrt{2}}$ prefactor, already takes into account that the  $\rho^0$-meson is described as $\frac{u\bar{u}-d\bar{d}}{\sqrt{2}}$.
 In total there are 20 diagrams which can contribute to both processes at twist-2 level. However in the case of  the $(\gamma\, \rho_T)$ production, due to the chiral-odd structure of the DA and of the GPD, only 8 diagrams do not vanish. By symmetry, it is enough to calculate 4 out of
those 8 diagrams, for example the four black diagrams of Figure~\ref{Fig:diagrams}.
In the case of   $(\gamma\, \rho_L )$ production, all 20 diagrams contribute, but, similarly as discussed above, by symmetry it is enough to calculate only 10 of them, shown in Figure~\ref{Fig:diagrams}. 

\section{RESULTS}

%
%
\psfrag{T}{} 
\def\sca{.7}
\psfrag{0}{\scalebox{\sca}{$0$}}
\psfrag{1}{\scalebox{\sca}{$1$}}
\psfrag{2}{\scalebox{\sca}{$2$}}
\psfrag{3}{\scalebox{\sca}{$3$}}
\psfrag{4}{\scalebox{\sca}{$4$}}
\psfrag{5}{\scalebox{\sca}{$5$}}
\psfrag{6}{\scalebox{\sca}{$6$}}
\psfrag{7}{\scalebox{\sca}{$7$}}
\psfrag{8}{\scalebox{\sca}{$8$}}
\psfrag{9}{\scalebox{\sca}{$9$}}
\psfrag{.}{\scalebox{\sca}{$.$}}
\begin{figure}[h!]
\psfrag{H}{\hspace{-1.5cm}\raisebox{-.6cm}{\scalebox{.7}{$M^2_{\gamma \rho}~({\rm GeV}^{2})$}}}
\psfrag{V}{\raisebox{.3cm}{\scalebox{.7}{$\hspace{-.4cm}\displaystyle\frac{d\sigma_{even}}{d M^2_{\gamma\rho}}~({\rm nb} \cdot {\rm GeV}^{-2})$}}}
\raisebox{-.13cm}{\includegraphics[width=6.8cm]{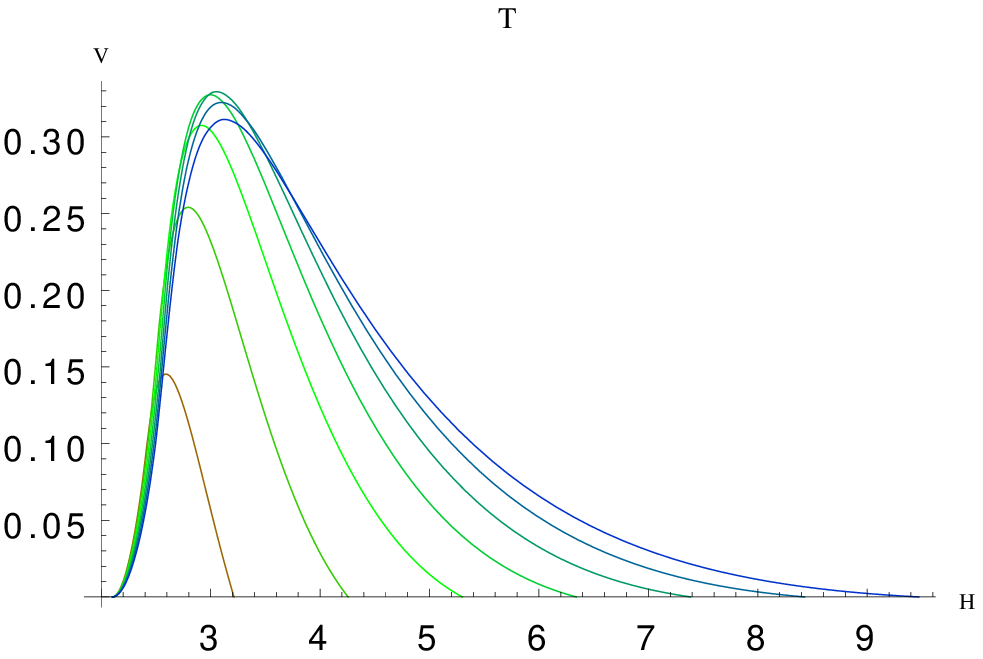}}
\psfrag{H}{\hspace{-1.5cm}\raisebox{-.6cm}{\scalebox{.7}{$M^2_{\gamma \rho} ({\rm GeV}^{2})$}}}
\psfrag{V}{\raisebox{.3cm}{\scalebox{.7}{$\hspace{-.7cm}\displaystyle\frac{d \sigma_{\rm odd}}{d M^2_{\gamma \rho}}~({\rm pb} \cdot {\rm GeV}^{-2})$}}}
\psfrag{T}{}
\raisebox{-.12cm}{\includegraphics[width=6.8cm]{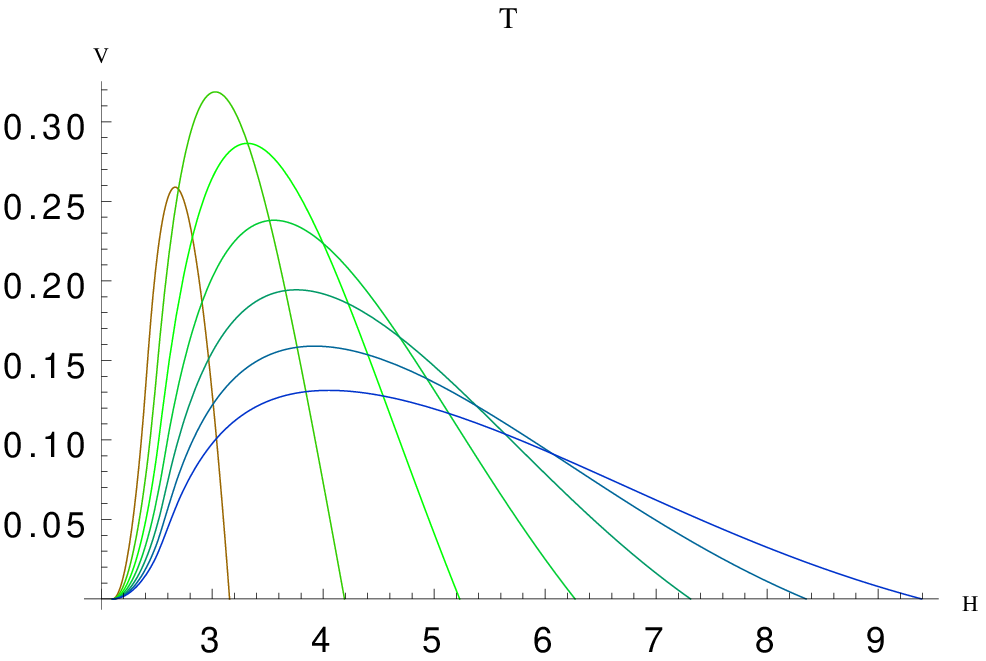}}
\vspace{.4cm}

\caption{Differential cross section $d\sigma/dM^2_{\gamma \rho}$ for a photon and a 
$\rho$ meson production on a proton target. The values of $S_{\gamma N}$ vary in the set 8, 10, 12, 14, 16, 18, 20 ${\rm GeV}^{2}.$ (from 8: left, brown to 20: right, blue), covering the JLab energy range. We use here the ``valence'' scenario.
Left: longitudinally polarized 
$\rho$. Right: transversally polarized 
$\rho$.
}
\label{Fig:dsigmaEVEN-dsigmaODD}
\end{figure}

\begin{figure}[!h]
\psfrag{T}{}
\includegraphics[width=6.2cm]{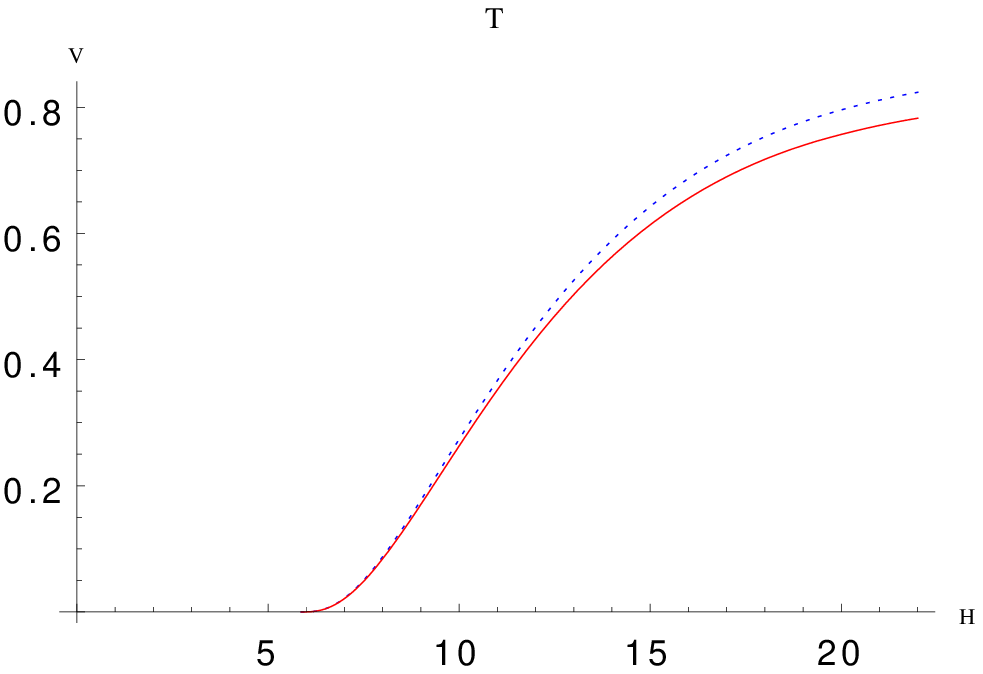}
\psfrag{H}{\hspace{-1.5cm}\raisebox{-.6cm}{\scalebox{.7}{$S_{\gamma N}~({\rm GeV}^{2})$}}}
\psfrag{V}{\raisebox{.3cm}{\scalebox{.7}{$\hspace{-.4cm}\displaystyle\sigma_{odd}~({\rm pb})$}}}
\qquad
\includegraphics[width=6.2cm]{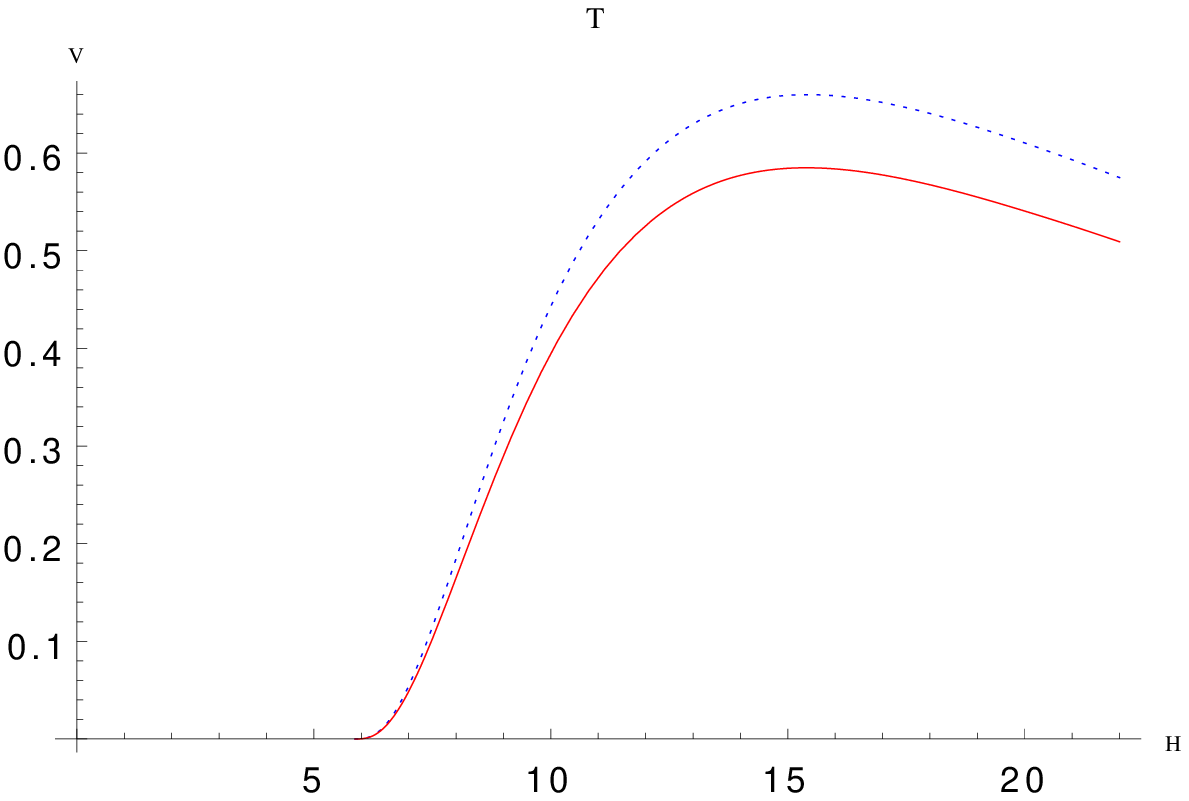}
\vspace{.4cm}

\caption{Integrated cross section for a photon and a   $\rho$ meson production on a proton target, as a function of $S_{\gamma N}.$ Left: longitudinally polarized $\rho$. Right:
transversally polarized $\rho$.
The solid red curves correspond to the ``valence'' scenario while the
dashed blue curves correspond to the ``standard'' one. 
}
\label{Fig:sigmaEVEN-sigmaODD}
\end{figure}

The differential cross-section 
$$
\left.\frac{d\sigma}{dt \,du' \, dM^2_{\gamma\rho}}\right|_{\ -t=(-t)_{min}} = \frac{|\mathcal{M}|^2}{32S_{\gamma N}^2M^2_{\gamma\rho}(2\pi)^3},
$$ 
a function of ($M^2_{\gamma\rho},u'$), 
is dominated by its chiral-even part.
Introducing a phenomenological $t$-dependence (a factorized dipole form in practice) and integrating over $t$ and $u'$ in a range consistent with the fixed angle hypothesis, we get the less differential cross section $\frac{d\sigma}{dM^2_{\gamma\rho}}.$ We model the GPDs via double distribution ans\"atze~\cite{Radyushkin:1998es}, based on unpolarized, polarized and transversity parton distributions. To get an order of magnitude of the uncertainties of this modeling, we use two types of polarized PDFs, hereafter named ``valence'' and ``standard''.
We show the differential cross-sections $\frac{d\sigma_{even}}{d M^2_{\gamma\rho}}$ and 
$\frac{d\sigma_{odd}}{d M^2_{\gamma\rho}}$ respectively
on the left (right) panel of
Figure~\ref{Fig:dsigmaEVEN-dsigmaODD},  
for various values of $S_{\gamma N}$ in the range accessible at JLab~12-GeV. The corresponding integrated cross sections
are displayed in Figure~\ref{Fig:sigmaEVEN-sigmaODD}.
The order of magnitude of the cross-section demonstrates that this experiment should be feasible at JLab~12-GeV, thus opening a new playground for the extraction of GPDs and a most useful test of their universality.
\psfrag{H}{\hspace{-1.5cm}\raisebox{-.6cm}{\scalebox{.7}{$S_{\gamma N} ({\rm GeV}^{2})$}}}
\psfrag{V}{\raisebox{.3cm}{\scalebox{.7}{$\hspace{-.4cm}\displaystyle\sigma_{even}~({\rm nb})$}}}

\section*{ACKNOWLEDGMENTS}

This work is partly supported by grant No 2015/17/B/ST2/01838 by the National Science Center in Poland, by the French grant ANR PARTONS (Grant No. ANR-12-MONU-0008-01), by the COPIN-IN2P3 agreement, by the Labex P2IO and by the Polish-French collaboration agreement Polonium.

%

\end{document}